\documentclass[twocolumn,showpacs,amsmath]{revtex4}
\usepackage{graphicx,amsmath,amsfonts}

\newcommand{\Journal}[4]{#1 \textbf{#2}, #3 (#4)}

\newcommand{\KLNO}{K$_{0.5}$Li$_{0.5}$NbO$_3$}
\newcommand{\KN}{KNbO$_3$}

\begin{document}

\title{Frustration of tilts and $A$-site driven ferroelectricity in KNbO$_3$-LiNbO$_3$ alloys}

\author{D. I. Bilc}
\author{D. J. Singh}
\affiliation{Condensed Matter Sciences Division, Oak Ridge National Laboratory, Oak Ridge, Tennessee 37831, USA, and Department of Physics and Astronomy, University of Tennessee, Knoxville, Tennessee 37996, USA}


\begin{abstract}

Density functional calculations for \KLNO\ show strong $A$-site driven ferroelectricity, even though the average tolerance factor is significantly smaller than unity and there is no stereochemically active $A$-site ion. This is due to the frustration of tilt instabilities by $A$-site disorder. There are very large off-centerings of the Li ions, which contribute strongly to the anisotropy between the tetragonal and rhombohedral ferroelectric states, yielding a tetragonal ground state even without strain coupling.

\end{abstract}

\maketitle


There is current interest in discovering new Pb free ferroelectric and piezoelectric materials for applications. (K,Na)NbO$_3$-LiTaO$_3$ ~\cite{Saito2004} textured materials were recently found to have similar piezoelectric performance to Pb(Zr,Ti)O$_3$ (PZT), which is the most used material for applications .~\cite{Jaffe1971, Uchino1996, Cross2004}  These systems are solid solutions with $AB$O$_3$ perovskite structure similar to PZT where the Pb atoms at the $A$-site are substituted by K, Na and Li atoms. This raises the question if a good room temperature Pb free relaxor ferroelectric material can be found. Also there is theoretical interest in microscopic understanding of the morphotropic phase boundary (MPB) between the rhombohedral and tetragonal ferroelectric phases, which plays a crucial role in the piezoelectric performance of materials.~\cite{Noheda1999, Guo2000, Fu2000}   

Lattice instabilities of perovskites are often discussed in terms of the tolerance factor, $t=(r_O+r_A)/2^{1/2}(r_O+r_B)$, where $r_O$, $r_A$, and $r_B$ are the ionic radii of the O, $A$, and $B$ ions.~\cite{Shannon1976,Zhong1995} Ferroelectrics with $t>$1 are called $B$-site driven, because the $B$-site ion is too small for its site and can off-center. These materials are typically ferroelectric with rhombohedral or similar structures and they do not show vertical MPB's. Examples of such materials are BaTiO$_3$, and \KN. Ferroelectrics with $t<$1 are called  $A$-site driven and they are in general not ferroelectrics because of different tilts and rotations of $B$O$_6$ octahedra, which preserve the inversion symmetry. Exceptions are materials with Pb$^{2+}$ as the $A$-site ion, as in PZT, for which the ferroelectricity occurs along with piezoelectrically active MPB's. The lattice instabilities can also be discussed in terms of relative off-centering of $A$- and $B$-site ions in their cages.~\cite{Ghita2005} This follows the classification in terms of $t$ except that ferroelectrics like PbTiO$_3$, with $t\sim$1 are classified as $A$-site driven due to Pb stereochemistry.

The piezoelectric performance depends on the ability of  switching between rhombohedral and tetragonal phase under an applied external field. This ability can be controlled by substitutions of the atoms at $A$- and $B$-site which favors one of the phase over the other. This has lead to several theoretical calculations in PZT systems, mainly investigating the $B$-site alloying. ~\cite{Saghi1999, Ramer2000, Fornari2001, Wu2003} Also recent studies of different $A$-site substitutions have showed increase in tetragonality when alloying with small $A$-site ions.~\cite{Halilov2002, Suares2003, Grinberg2003, Halilov2004} 

Here, we use density functional calculations to explore the lattice instabilities of ordered supercells of \KLNO\ with perovskite structure. Since there are not stereochemically active ions like Pb$^{2+}$ or Bi$^{3+}$, one does not anticipate strong $A$-site driven ferroelectricity. However we find it due to a new mechanism-frustration of tilts by $A$-site disorder.\cite{Halilov2002,Halilov2004}  There are not experimental results for the 50$\%$/50$\%$ composition of K/Li at the $A$-site probably due to difficulty to synthesize \KLNO\ in perovskite structure. (K,Li)NbO$_3$ around 50$\%$/50$\%$ occupancy does not normally occur in perovskite structure, but rather occurs in tungsten bronze structure in a fairly broad composition range around 60$\%$ K. However, the perovskite structure is a higher density phase and should be accessible with appropriate growth techniques, perhaps high pressure methods.


The calculations were performed using density functional theory within local density approximation (LDA). Although the energy differences discussed here are small, the LDA is known to correctly describe ferroelectricity in perovskites. Specifically, the end point compounds, KNbO$_3$ ~\cite{Singh1992, Singh1996} and LiNbO$_3$~\cite{Inbar1996} are well described and even delicate balances like the  non-ferroelectric character of KTaO$_3$~\cite{Singh1996} are reproduced.
 We use the linearized augmented plane wave method (LAPW) ~\cite{Singh1994} as implemented in both WIEN2K~\cite{WIEN2K} and our in-house code, including local orbitals ~\cite{Singh1991} to relax liniarization errors and to treat the high lying semicore states of the metal ions. The values of the lapw sphere radii were taken small for the relaxations. These values are: 2.0 a.u. for K, 1.85 a.u. for Li and Nb and  1.5 a.u. for O. Convergence of the self-consistent iterations was performed for a mesh of (6,6,6) or equivalent grids of  $\vec{k}$ points inside the Brillouin  zone to within 0.0001 Ry with a cutoff of -6.0 Ry between the valence and the core states. The number of plane waves used in the interstitial region is characterized by a parameter RK$_{max}$=R$_{ml}$K$_{max}$, where R$_{ml}$ is the O sphere radius and K$_{max}$ is the maximum plane wave vector. For our calculations RK$_{max}$ was taken to be 8. The core states were treated fully relativistically, whereas the valence states were treated within a scalar relativistic approximation. The calculations for the \KLNO\ were done using 2x2x2 fcc supercells with 10 atoms/cell and also using different orderings of K and Li atoms in 2x2x2 sc supercells with 40 atoms/cell. We use the theoretical lattice constant (3.934\AA) since the system considered here has not been synthesized experimentally.


In order to understand the lattice distortions and the trends in the ferroelectric instabilities we have considered these instabilities separately in the 10 atom supercell. We performed LDA calculations at the theoretical volume for rhombohedral and tetragonal ferroelectric states, as well as octahedral rotations around [001] and [111] cubic axis. The calculations were carried out by imposing space group symmetries that allow the desired mode (centrosymmetric with 10 atoms/cell for the rotations and noncentrosymmetric with primitive cells for the ferroelectric states, which allow the ferroelectric displacements, but not octahedral rotations). For the primitive perovskite cell the octahedral rotation is a pure mode at the $R$ point ($R_{25}$), while the ferroelectric modes are part of a manifold of transverse optic modes. We calculated the energetics for the octahedral rotations around the two axes ([001] and [111]), whereas for the ferroelectric modes we have fully relaxed the atomic positions in the unit cells. In all the calculations, except the tetragonal ferroelectric, the lattice parameters were taken at the theoretical cubic values. For the tetragonal ferroelectric state we have varied the $c/a$ ratio keeping the unit cell volume constant.

\begin{figure}[t]
   \centering\includegraphics[scale=0.30, angle=-90]{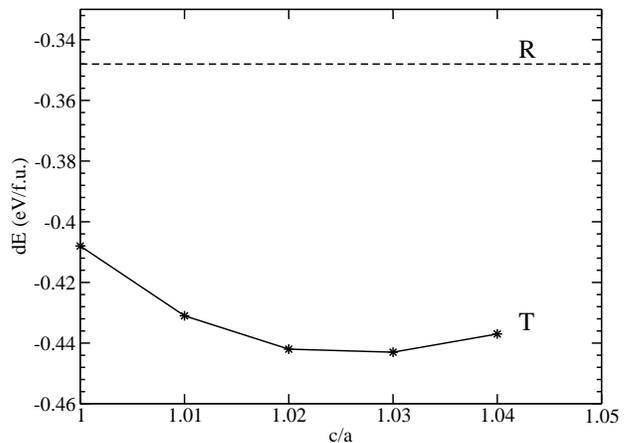}
    \caption{\label{strainfig} Total energy for the strained tetragonal (T) ferroelectric state of \KLNO\ fcc ordered supercell relative to the undistorted cubic structure as a function of the c/a ratio. The energy of the rhombohedral (R) ferroelectric state is shown in dashed line.}
\end{figure}

The total energy calculations for the different instabilities give a tetragonal ferroelectric ground state with the two rotational distortions being strongly suppressed. Even without the strain the tetragonal ferroelectric state is the ground state. This is different from the other known perovskite materials. The energies relative to the undistorted cubic structure for the fully relaxed rhombohedral (R) and tetragonal (T) ferroelectric, octahedrally rotated around [111] (R111), and octahedrally rotated around [001] (R001) structures are: -0.348eV/f.u., -0.408eV/f.u., -0.023eV/f.u., and -0.020eV/f.u. (f.u.=2\KLNO) respectively. The strain calculations for the tetragonal ferroelectric state lower the energy even more giving the lowest value of -0.443eV/f.u. at 3$\%$ strain. These results are shown in Fig.~\ref{strainfig}. The calculations show a energy difference of $\sim$95meV/f.u. between the strained tetragonal and rhombohedral ferroelectric states suggesting the possible presence of an active MPB, between this composition and KNbO$_3$ which is rhombohedral.

\begin{figure}[t]
   \centering\includegraphics[scale=0.30, angle=-90]{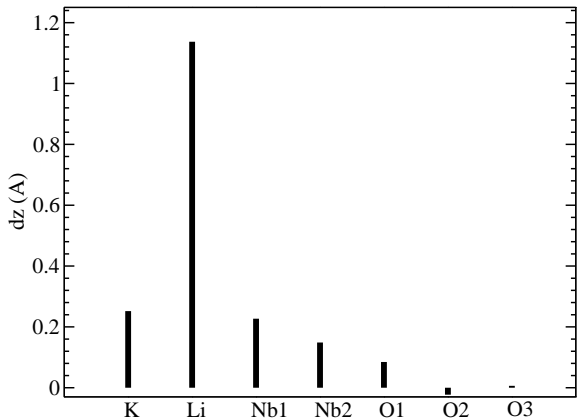}
    \caption{\label{displacementfig} Atom displacements relative to the center of mass of the O atoms for the 3$\%$ strained tetragonal (T) ferroelectric state of \KLNO\ fcc ordered supercell.}
\end{figure}

The ferroelectric distortions in the perovskite materials consist of displacements of the cations ($A$- and $B$-site ions) relative to the center of mass of the surrounding O anions. In order to see which cation displacements are the driving force to stabilize the tetragonal ferroelectric structure, we have plotted the atomic displacements of all the atoms relative to the O's center of mass for the lowest energy tetragonal structure (3$\%$ strained tetragonal structure) along the [001] direction. These displacements are given in Fig.~\ref{displacementfig}. The Li ions have very large displacements. Relative to this, the Nb and K displacements are small. This is in contrast to most perovskite ferroelectrics, where there is significant off-centering of all ions. The tetragonal ferroelectric state occurs as a consequence of large Li displacements. This result can be understood in terms of frustration arising from the very different ionic radii of K and Li. Since the NbO$_6$ octahedra are relatively stiff, the coherence length of the octahedral rotations is much longer than that of $A$-site off-centering.~\cite{Halilov2002,Halilov2004, Ghosez1999, Yu1995} Because of this, the frustration of K and Li ions favors the $A$-site off-centering over the octahedral rotations. Information about the coherence length can be obtained from known phonon dispersions. In cubic KNbO$_3$~\cite{Yu1995} sheets of instabilities are found whereas cubic PbTiO$_3$~\cite{Ghosez1999} has an unstable mode over the whole zone. In general, tilt instabilities exist only near the zone boundary and therefore have a long coherence length. Also the small values of critical thickness in PbTiO$_3$ thin films show the small coherence length of the ferroelectric state in A-site driven materials.~\cite{Ghosez2000}

\begin{figure}[t]
   \centering\includegraphics[scale=0.35, angle=-90]{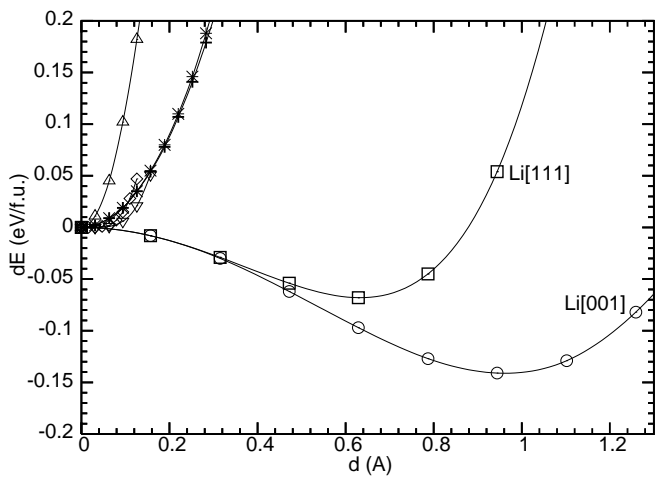}
    \caption{\label{individdisplfig} Energy vs individual atomic displacements around T [001] and R [111] directions of \KLNO\ fcc ordered supercell. The values are shown in: $\circ$ for Li along T, $\square$ for Li along R, $\Diamond$ for both Nb atoms along T, $\bigtriangledown$ for both Nb atoms along R, $\bigtriangleup$ for one Nb atom along R, $+$ for K along T, and $\star$ for K along R directions.}
\end{figure}

In order to see more clearly the individual atomic contributions to the tetragonal and rhomohedral ferroelectric states all the cations were individually displaced around the tetragonal [001] and rhombohedral [111] equilibria while keeping all the other atoms fixed. These results are given in Fig.~\ref{individdisplfig} and they clearly show that Li offcenters and favors the [001] direction over [111] by 73meV/(f.u.). Because of the large Li off-centering, Li contributes strongly to the anisotropy between the tetragonal and rhombohedral ground states, yielding a tetragonal ground state. This is in contrast to the usual case in perovskite ferroelectrics where strain coupling may stabilize the tetragonal state, but otherwise the anisotropy comes from the $B$-site and favors rhombohedral states. This is significant because having a material with a true tetragonal state is important for obtaining a vertical MPB when alloying with a rhombohedral material, as is the case in PZT. Individual displacements of Nb and K atoms do not show any instability toward ferroelectric states.

\begin{figure}[t]
   \centering\includegraphics[scale=0.18]{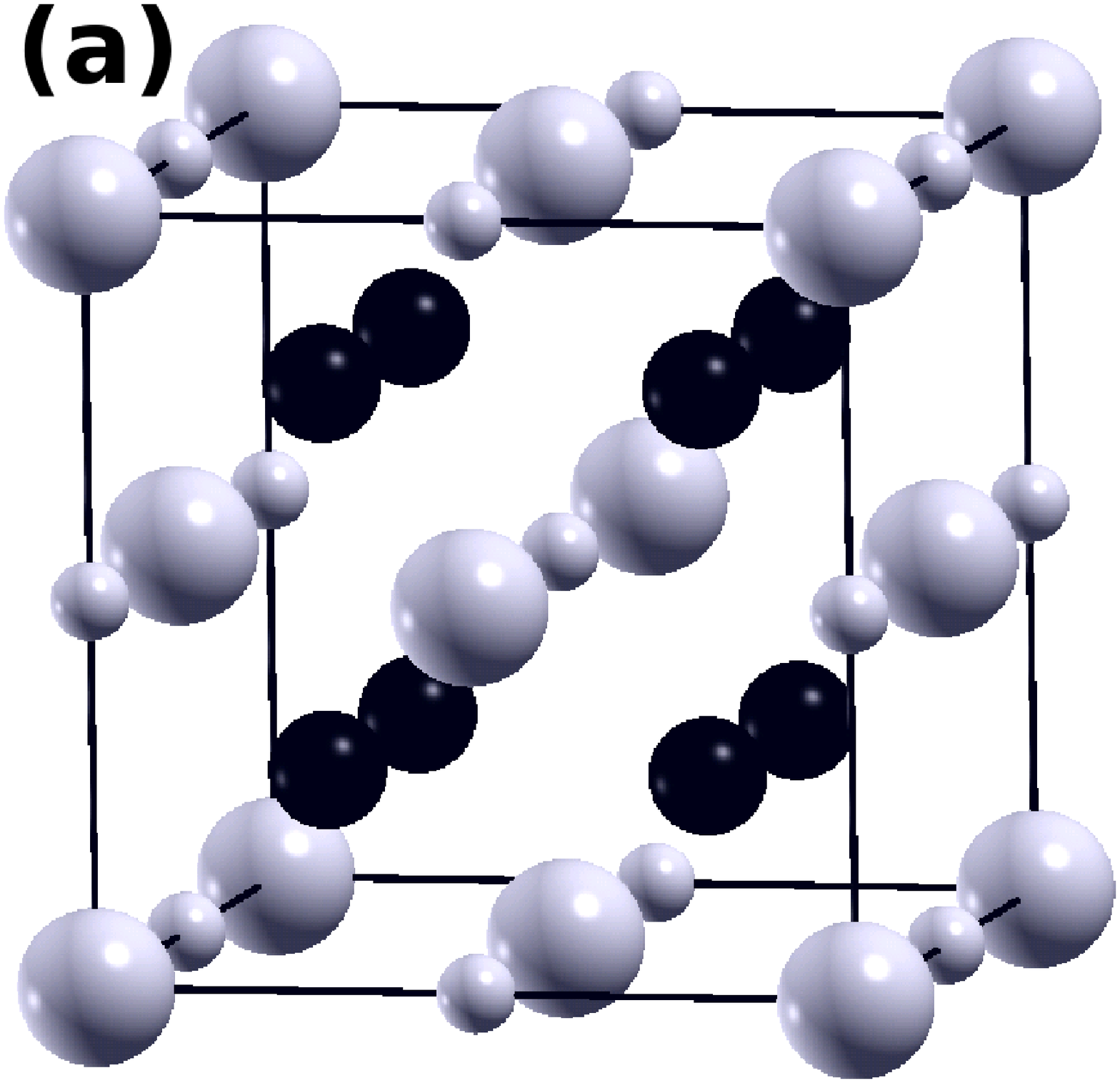} %
   \centering\includegraphics[scale=0.18]{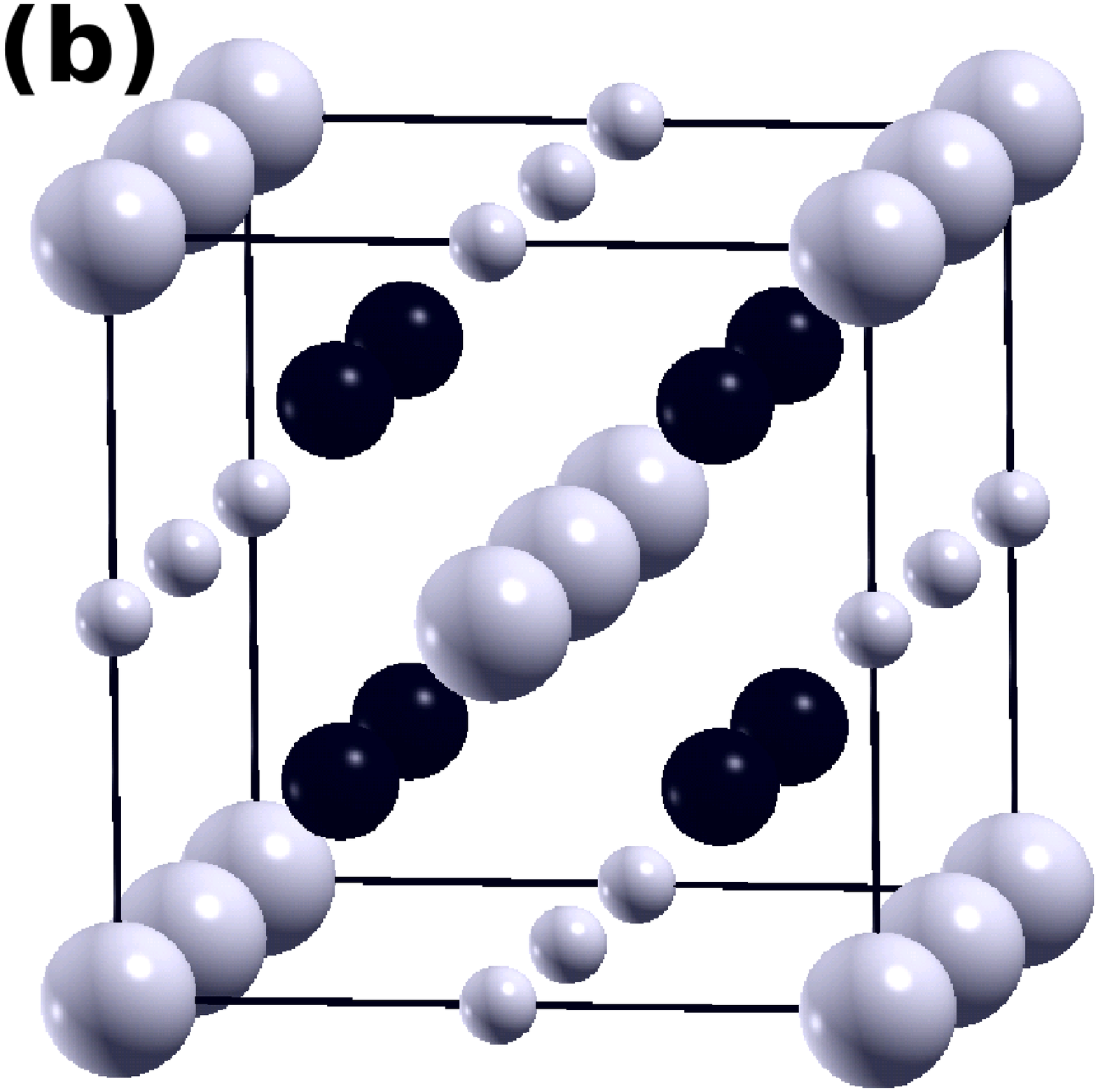}
   \centering\includegraphics[scale=0.185]{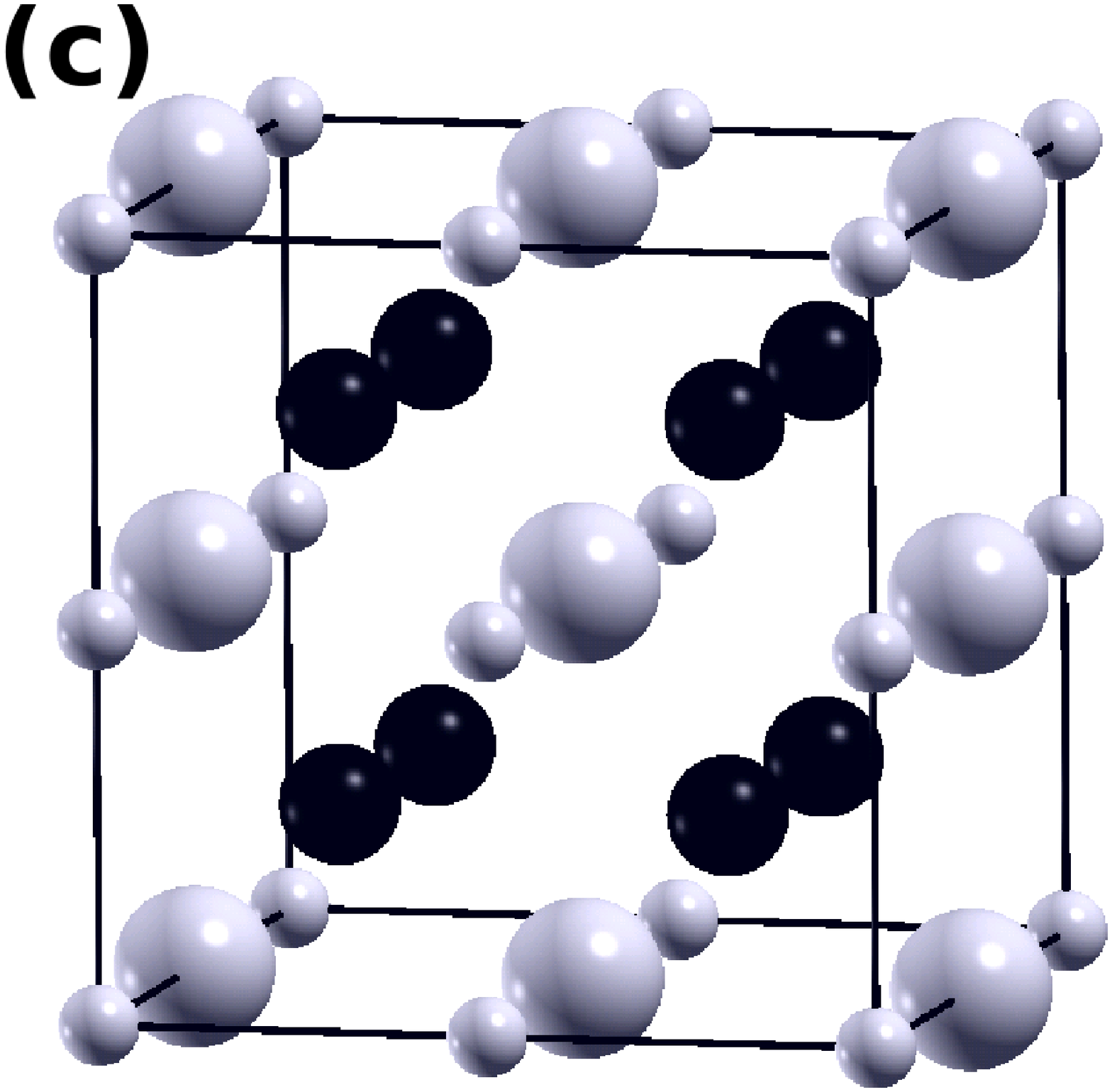} %
   \centering\includegraphics[scale=0.18]{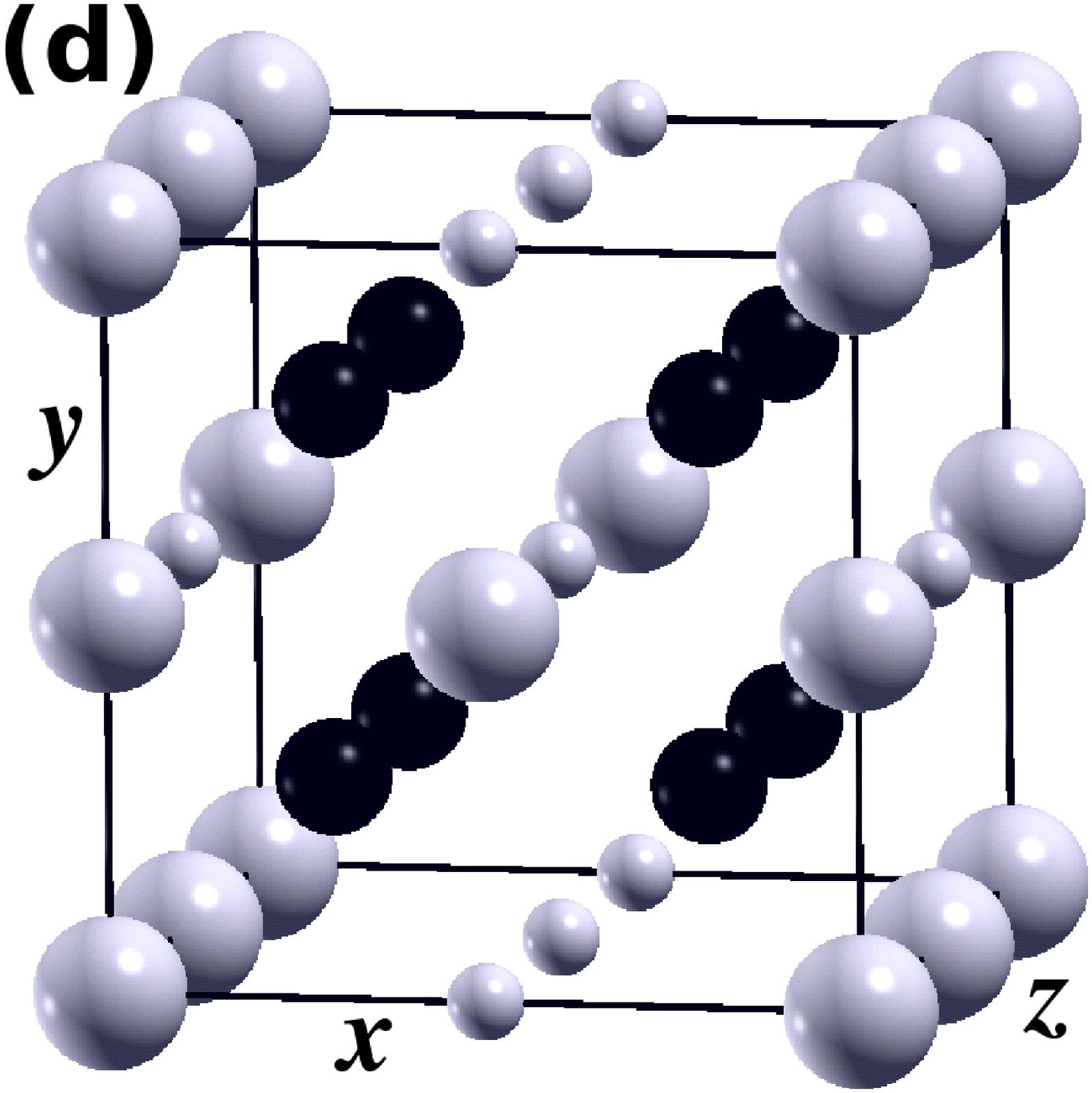}
    \caption{\label{scunitcellfig} Unit cells for the 40 atom supercells with different orders: (a) fcc order, (b) chain order along z-axis, (c) layer order perpendicular to the z-axis, and (d) the most disordered case for this type of supercell. The K, Li and Nb ions are shown in large gray, small gray, and black colors, respectively. For the reason of clarity O anions are not shown.}
\end{figure}

To see the effect of different K and Li arrangements on the ferroelectric state we performed full relaxations on different K/Li ordered 2x2x2 supercells with 40 atoms/cell. We have considered four different K/Li ordered supercells : fcc order, chain order along the z-axis, layered perpendicular to the z-axis, and a disordered case for this type of supercell. These are shown in Fig.~\ref{scunitcellfig}. The total energies of the fully relaxed cells relative to that of the supercell with fcc order are: 0meV/(f.u.) for fcc order, -107.12meV/(f.u.) for chain order, +88.54meV/(f.u.) for layer order, and -17.7meV/(f.u.) for the most disordered case. The most stable ordered configuration is the chain order. However, if this material follows the pattern of other perovskite oxides, samples will be disordered as the cations become mobile only at high temperature. The supercells give ferroelectric states and not octahedral rotations as the ground states with Li ions having the largest displacements. For the fcc order and chain order supercells, Li displacements are along the tetragonal [001] direction. For the chain order the Li displacements ($\sim$1.5\AA) are larger that those of the fcc order, whereas the K displacements are smaller. The chain order better minimizes the strain by larger Li displacements yielding the lowest energy. 

\begin{figure}[t]
  \begin{minipage}[h]{0.5\textwidth}
  \begin{minipage}[b]{0.5\textwidth}
   \centering\includegraphics[scale=0.60]{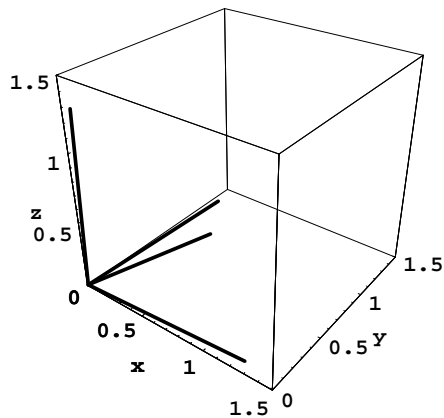}
  \end{minipage}\\[-20pt]
    \caption{\label{disordercationdisplacementfig} Li displacements relative to the center of mass of the O atoms for the disordered supercell.}
  \end{minipage}
\end{figure}

The most disordered case (Fig.~\ref{scunitcellfig}d) for which the 3D plots of the Li displacements are shown in Fig~\ref{disordercationdisplacementfig} is particularly interesting. Three of the Li ions off-center almost along the three tetragonal directions and one Li ion off-centers to a more general direction given by the displacement vector [0.801,0.380,0.467] (Fig~\ref{disordercationdisplacementfig}). The direction of movement for the three Li which displace close to the tetragonal directions is toward the next nearest neighbor Li ions. The Li displacements show frustration of the Li in finding the optimal direction of off-centering. Such a property is favorable for forming a relaxor ferroelectric.


To summarize, lattice instabilities of supercells of \KLNO\ were investigated using electronic structure calculations. Frustration between large K and small Li atoms at the $A$-site leads to strong suppression of the rotational instabilities and the presence of $A$-site driven ferroelectricity with large Li off-centering. This favors the tetragonal ferroelectric ground state even without strain. This is different than the other perovskite materials which have a rhomohedral ferroelectric state without strain inclusion.

$A$-site driven ferroelectrics like PZT also have ferroelectric coherence lengths shorter than the $B$-site driven materials.~\cite{Ghosez1999} This and the coupling of Li off-centering to the local Li-K ordering suggest that materials near this composition may be good relaxor ferroelectrics.

\KLNO\ system is an example of alloying large and small $A$-site cations in order to create strong $A$-site ferroelectricity with a tetragonal ferroelectric ground state. This is significant because it suggests ways to find Pb free perovskites with nearly vertical MPB's.

\begin{acknowledgments}
 
We are grateful for helpful discussions with P. K. Davies, T. Egami, M. Fornari, S. V. Halilov, and M. Suewattana. Research sponsored by the Division of Materials Sciences and Engineering, Office of Basic Energy Sciences, U.S. Department of Energy, under contract DE-AC05-00OR22725 with Oak Ridge National Laboratory, managed and operated by UT-Battelle, LLC, and the Office of Naval Research.

\end{acknowledgments}

\end{document}